\documentclass{sig-alternate}
\usepackage{amsmath,latexsym,amssymb,verbatim}
\usepackage[multiple]{footmisc}

\def\bR{{\mathbb R}}

\def\CA{{\mathcal A}}

\def\Mfont#1{\mathsf{#1}}
\def\ALL{{\Mfont{ALL}}}
\def\LO{{\Mfont{LO}}}
\def\LOCC{{\Mfont{LOCC}}}

\def\leftLOCC{{ \Mfont{LOCC}^\leftarrow }}
\def\leftrightLOCC{{{\Mfont{LOCC}}}}
\def\Mclass{{\Mfont{M}}}
\def\SEP{{\Mfont{SEP}}}
\def\CS{{\mathcal{S}}}

\def\QMA{{\sf{QMA}}}

\def\BellQMA{{\sf{BellQMA}}}

\def\NP{{\sf{NP}}}

\def\MA{{\sf{MA}}}

\def\P{{\sf{P}}}

\def\SAT{{\sf{SAT}}}

\def\nn{\nonumber}
\def\norm#1{ {\Vert #1\Vert} }
\def\Norm#1{ {\big\Vert #1 \big\Vert} }

\def\squareforqed{\hbox{\rlap{$\sqcap$}$\sqcup$}}
\def\qed{\ifmmode\squareforqed\else{\unskip\nobreak\hfil
\penalty50\hskip1em\null\nobreak\hfil\squareforqed
\parfillskip=0pt\finalhyphendemerits=0\endgraf}\fi}
\def\endenv{\ifmmode\;\else{\unskip\nobreak\hfil
\penalty50\hskip1em\null\nobreak\hfil\;
\parfillskip=0pt\finalhyphendemerits=0\endgraf}\fi}

\newcommand{\proj}[1]{|#1\rangle\langle #1|}
\newcommand{\bra}[1]{\langle #1|}
\newcommand{\ket}[1]{|#1\rangle}

\DeclareMathOperator{\tr}{tr}

\newcommand{\id}{\mathbb{I}}

\mathchardef\ordinarycolon\mathcode`\:
\mathcode`\:=\string"8000
\def\vcentcolon{\mathrel{\mathop\ordinarycolon}}
\begingroup \catcode`\:=\active
  \lowercase{\endgroup
  \let :\vcentcolon
  }

\newcommand{\nc}{\newcommand}
\nc{\rnc}{\renewcommand} \nc{\beq}{\begin{equation}}
\nc{\eeq}{{\end{equation}}} \nc{\bea}{\begin{eqnarray}}
\nc{\eea}{\end{eqnarray}} \nc{\beqa}{\begin{eqnarray}}
\nc{\eeqa}{\end{eqnarray}} \nc{\lbar}[1]{\overline{#1}}

\nc{\conv}{\operatorname{conv}}
\nc{\smfrac}[2]{\mbox{$\frac{#1}{#2}$}} \nc{\Tr}{\operatorname{Tr}}
\nc{\ox}{\otimes} \nc{\dg}{\dagger} \nc{\dn}{\downarrow}
\nc{\lmax}{\lambda_{\text{max}}}
\nc{\lmin}{\lambda_{\text{min}}}

\nc{\csupp}{{\operatorname{csupp}}}
\nc{\qsupp}{{\operatorname{qsupp}}} \nc{\var}{\operatorname{var}}
\nc{\rar}{\rightarrow} \nc{\lrar}{\longrightarrow}
\nc{\poly}{\operatorname{poly}}
\nc{\polylog}{\operatorname{polylog}} \nc{\Lip}{\operatorname{Lip}}
\nc{\mb}[1]{\mathbf{#1}}
\nc{\ep}{\epsilon}
\nc{\Om}{\Omega}
\nc{\wt}[1]{\widetilde{#1}}

\def\>{\rangle}
\def\<{\langle}

\def\x{\xi}

\def\S{\Sigma}

\nc{\glneq}{{\raisebox{0.6ex}{$>$}  \hspace*{-1.8ex} \raisebox{-0.6ex}{$<$}}}
\nc{\gleq}{{\raisebox{0.6ex}{$\geq$}\hspace*{-1.8ex} \raisebox{-0.6ex}{$\leq$}}}

\nc{\vholder}[1]{\rule{0pt}{#1}}
\nc{\wh}[1]{\widehat{#1}}
\nc{\h}[1]{\widehat{#1}}

\nc{\ob}[1]{#1}

\def\beq{\begin {equation}}
\def\eeq{\end {equation}}

\def\be{\begin{equation}}
\def\ee{\end{equation}}

\nc{\eq}[1]{Eq.~(\ref{eq:#1})} \nc{\eqs}[2]{Eqs.~(\ref{eq:#1}) and
(\ref{eq:#2})}

\nc{\eqn}[1]{Eq.~(\ref{eqn:#1})}
\nc{\eqns}[2]{Eqs.~(\ref{eqn:#1}) and (\ref{eqn:#2})}

\nc{\region}{\cS\cW}

\newtheorem{theorem}{Theorem}
\newtheorem{corollary}{Corollary}
\newtheorem{definition}{Definition}

\newtheorem{lemma}{Lemma}
\newtheorem{problem}{Problem}

\begin{document}

%

\title{A Quasipolynomial-Time Algorithm \\ for the Quantum Separability Problem}

\numberofauthors{3} 
\author{
\alignauthor
Fernando~G.S.L.~Brand\~ao\titlenote{Departamento de F\'isica, Universidade Federal de Minas Gerais, Belo Horizonte, Brazil.  Supported by a ``Conhecimento Novo" fellowship from FAPEMIG.\\
	{\tt fgslbrandao@gmail.com}} \\ 
	\affaddr{\vspace{-0.1in} Universidade Federal de Minas Gerais}\\
\alignauthor
Matthias Christandl\titlenote{Institute for Theoretical Physics, ETH Zurich, Wolfgang-Pauli-Strasse 27, CH-8057 Zurich, Switzerland.  Supported by the Swiss National Science Foundation (grant PP00P2-128455) and the German Science Foundation (grants~CH 843/1-1 and CH 843/2-1). {\tt christandl@phys.ethz.ch}} \\ 
	\affaddr{\vspace{.045in} ETH Zurich}
\alignauthor
Jon Yard\titlenote{Center for Nonlinear Studies (CNLS) \& Computer, Computational and Statistical Sciences (CCS-3), Los Alamos National Laboratory, Los Alamos, NM 87545. Supported by grants through the LDRD program of the US Department of Energy.  {\tt jtyard@lanl.gov}} \\ 
	\affaddr{\vspace{.045in} Los~Alamos~National Laboratory}\\
}

\maketitle
\begin{abstract}
We present a quasipolynomial-time algorithm for solving the weak membership problem for the convex set of separable, i.e.\ non-entangled, bipartite density matrices.  The algorithm decides whether a density matrix is separable or whether it is  $\epsilon$-away from the set of the separable states in time $\exp(O(\epsilon^{-2}\log|A| \log|B|)),$ where $|A|$ and $|B|$ are the local dimensions, and the distance is measured with either the Euclidean norm, or with the so-called LOCC norm.  The latter is an operationally motivated norm giving the optimal probability of distinguishing two bipartite quantum states, each shared by two parties, using any protocol formed by quantum local operations and classical communication (LOCC) between the parties. We also obtain improved algorithms for optimizing over the set of separable states and for computing the ground-state energy of mean-field Hamiltonians.

The techniques we develop are also applied to quantum Merlin-Arthur games, where we show that multiple provers are not more powerful than a single prover when the verifier is restricted to LOCC protocols, or when the verification procedure is formed by a measurement of small Euclidean norm. This answers a question posed by Aaronson \emph{et al.} (Theory of Computing {\bf 5}, 1, 2009) and provides two new characterizations of the complexity class ${\sf QMA}$, a quantum analog of ${\sf NP}$. 

Our algorithm uses semidefinite programming to search for a symmetric extension, as first proposed by Doherty, Parrilo and Spedialieri (Phys. Rev. A, 69, 022308, 2004). The bound on the runtime follows from an improved de Finetti-type bound quantifying the monogamy of quantum entanglement. This result, in turn, follows from a new lower bound on the quantum conditional mutual information and the entanglement measure squashed entanglement.
\end{abstract}



\section{Introduction}
A central problem in quantum information theory is to characterize entanglement in quantum states shared by two or more parties \cite{HHHH09}.  A bipartite density matrix, or \emph{state}, is a positive semidefinite matrix $\rho_{AB}$ on the tensor product $AB \equiv A\ox B$ of finite dimensional complex vector  spaces that is \emph{normalized}, meaning $\tr(\rho_{AB}) = 1$.  Such a state is \textit{separable} if it can be written as $\rho_{AB} = \sum_{k}p_k \rho_{A, k} \otimes \rho_{B, k}$, for local states $\rho_{A, k}$ and $\rho_{B, k}$ and probabilities $p_k$. Any separable state can be created by local quantum operations and classical communication (LOCC) by Alice and Bob and thus only contains classical correlations. Quantum states that are not separable are called \textit{entangled}.  As the normalized Hermitian matrices on $AB$ form a real vector space of dimension $d = |A|^2 |B|^2 -1$ (we abbreviate $\dim(A) = |A|$), the set of all states can be viewed as a compact, convex subset of $\bR^d$ containing the convex subset $\CS \equiv \CS_{A:B}$ of separable states.  

A fundamental question is to decide, given a description of $\rho_{AB}$ (say, as a rational vector in $\bR^d$) whether or not it is separable \cite{HHHH09, DPS04, Ioa07, Gur03, Gha10, HM10}, i.e.\ whether or not it is contained in $\CS$. This can be formalized as a decision problem via the weak membership problem.
Given a norm $\norm{*}$ on $\bR^d$ and a closed subset $\CA \subset \bR^d$, let $\norm{\rho - \CA} = \min_{\sigma \in \CA} \norm{\rho - \sigma}$ be the distance from $\rho$ to $\CA$.
\begin{problem}
\textsc{$W_{\text{SEP}}(\epsilon, \Vert * \Vert)$ (Weak membership problem for separability):} Given a density matrix $\rho_{AB}$ with the promise that either (i) $\rho_{AB} \in \CS$ or (ii) $\norm{\rho_{AB}-\CS} \geq \ep$, decide which is the case. 
\end{problem}

This problem has been intensely studied in recent years (see e.g.\  \cite{HHHH09, DPS04, Ioa07, Gur03, Gha10, HM10}) with the norm given either by the Euclidean norm $\Vert X \Vert_2 \equiv \tr(X^{\cal y}X)^{1/2}$ or by the trace norm $\Vert X \Vert_{1} \equiv \tr \sqrt{X^\dagger X}$. 

The best-known algorithms for $W_{\text{SEP}}(\epsilon, \norm{*})$ \cite{DPS04, Ioa07} (with the norm equal either to Euclidean or trace norm) have worst-case complexity $\exp \left( O \left( |A|^{2}|B|^{2} \log(\epsilon^{-1}))  \right) \right)$. On the hardness side, Gurvits \cite{Gur03} proved that $W_{\text{SEP}}(\epsilon, \Vert * \Vert_2)$ is $\NP$-hard for $\epsilon = \exp(-O(d))$, with $d = \sqrt{|A|\cdot |B|}$; the dependence on $\epsilon$ was later improved to $\epsilon = 1/\poly(d)$ \cite{Gha10}. The same results apply to the trace norm, since for every $l \times l$ matrix, $\Vert X \Vert_1 \geq \Vert X \Vert_2 \geq l^{-1/2} \Vert X \Vert_1$. 

A second problem closely related to the weak-membership problem for separability is the following:
\begin{problem}
\textsc{BSS($\epsilon$) (Best Separable State):} Given a Hermitian matrix  $M$ on $A B$, estimate $\max_{\sigma \in {\cal S}} \tr(M \sigma)$ with additive error $\epsilon$.
\end{problem}

The \textsc{BSS($\epsilon$)} problem thus consists of optimizing a linear function over the convex set of separable states ${\cal S}$. It is a standard fact in convex optimization \cite{GLS93} that linear optimization and weak-membership over a convex set are equivalent tasks, which implies that \textsc{BSS($\epsilon$)}  can be used to solve $W_{\text{SEP}}(\delta, \Vert * \Vert)$ and vice-versa, up to a $\poly(d)$ loss in the error parameters $\epsilon$ and $\delta$ (see~\cite{Ioa07} for a detailed analysis). The best known algorithm for \textsc{BSS($\epsilon$)} has worst-case complexity $\exp \left(O\left( (|A|+|B|) \Vert M \Vert_{\infty} \log(\epsilon^{-1}) \right) \right)$.\footnote{$\Vert M \Vert_{\infty}$ is the \textit{operator norm} of $M$, given by the maximum eigenvalue of $\sqrt{M^{\cal y}M}$.}\footnote{The algorithm is nothing more than exhaustive search, in which for an operator $M$ with $\Vert M \Vert_{\infty} \leq 1$, one considers $\epsilon$-nets~\cite[Lemma II.4]{HLSW04} $\{  \ket{a_k} \}_k$ and $\{ \ket{b_k} \}_j$ for the $A$ and $B$ systems of sizes $\exp \left( O(|A| \log(\epsilon^{-1})) \right)$ and $\exp \left( O(|B| \log(\epsilon^{-1})) \right)$, respectively, and minimize $\bra{a_k, b_j} M \ket{a_k, b_j}$ over $k, j$.} The $\NP$-hardness of the weak-membership problem for separability implies that \textsc{BSS($\epsilon$)} is $\NP$-hard for $\epsilon = 1/\poly(d)$. Conditioned on the stronger assumption that there is no subexponential-time algorithm for 3-$\SAT$ \cite{IP01}, Harrow and Montanaro \cite{HM10}, building on work by Aaronson et al.\ \cite{ABDFS08}, recently ruled out even quasipolynomial-time algorithms for \textsc{BSS($\epsilon$)} of complexity up to $\exp \left( O \left(\log^{1 - \nu} |A|    \log^{1 - \mu} |B| \Vert M \Vert_{\infty}\right) \right)$ for \textit{constant} $\epsilon$ and any $\nu + \mu > 0$. More specifically, they showed one could solve 3-$\SAT$ with $n$ clauses by solving \textsc{BSS($\epsilon$)}, with constant $\epsilon$, for a matrix $0 \leq M \leq \id$ on $AB$ with $|A| = |B| = 2^{O(\sqrt{n}\poly\log(n))}$. Indeed, this shows that an algorithm for \textsc{BSS($\epsilon$)} with time complexity 
\begin{eqnarray*}
\exp \left( O \left(\log^{1 - \nu} |A|    \log^{1 - \mu} |B| \Vert M \Vert_{\infty}\right) \right) 
\end{eqnarray*}
 would imply an $\exp \left( O( n^{1 - (\nu + \mu)/2}\poly\log(n)) \right)$-time algorithm for 3-$\SAT$.

The best separable state problem has a number of other applications (see e.g.\ \cite{HM10}), including the estimation of the ground-state energy of mean-field quantum Hamiltonians and estimating the minimal min-entropy of quantum channels. In entanglement theory, it has been studied under the name of optimization of entanglement witnesses (see e.g.\ \cite{HHHH09}).\footnote{An entanglement witness $W$ is a Hermitian operator which has positive trace on all separable states and a negative trace on a particular entangled state, thus witnessing the fact that the state is entangled.}

It turns out that the problem \textsc{BSS($\epsilon$)} is also intimately connected to quantum Merlin-Arthur games with multiple Merlins. The class $\QMA$ is a quantum analog of $\NP$ and is formed by all languages that can be decided in quantum polynomial-time by a verifier who is given a quantum system of polynomially many qubits as a proof (see e.g.\ \cite{Wat08}). The class $\QMA(2)$, in turn, is a variant of $\QMA$ in which two proofs, not entangled with one another, are given to the verifier \cite{KMY03}. The properties of $\QMA(2)$ and its relation to $\QMA$ have recently been in the center of interest in quantum complexity theory \cite{HM10, KMY03, ABDFS08, BT07, Wat08, Bra08, Liu, Bei08, CD10}. As shown in \cite{HM10}, the optimal acceptance probability of a $\QMA(2)$ protocol can be expressed as a \textsc{BSS($\epsilon$)} instance. Thus a better understanding of the latter would also shed light on the properties of $\QMA(2)$.

\section{Results}

\vspace{0.2 cm}
\noindent \textbf{A quasipolynomial-time algorithm for separability:} Our main result is a quasipolynomial-time algorithm for $W_{\text{SEP}}(\epsilon, \Vert * \Vert)$, for two different choices of the norm: 
\begin{theorem} \label{main}
$W_{\rm{SEP}}(\epsilon, \Vert * \Vert_2)$ and $W_{\rm{SEP}}(\epsilon, \Vert * \Vert_{\Mfont{LOCC}})$ 
can be solved in  $\exp\!\left( O(\ep^{-2}\log|A| \log|B|) \right)$ time.
\end{theorem}

The norm $\norm{*}_\LOCC$ can be seen as a restricted version of the trace norm $\norm{*}_1 $. The latter can be written as
$$\norm{X}_1= \max_{0 \leq M \leq \id} \tr((2M - \id) X),$$ where $\id$ is the identity matrix, and is of special importance in quantum information theory as it is directly related to the optimal probability for distinguishing two equiprobable states $\rho$ and $\sigma$ with a quantum measurement.\footnote{Two-outcome measurements suffice for such tasks, and these are described by a pair of positive semidefinite matrices summing to $\id$, which we write $\{M,\id - M\}$.  When the state is $\rho$, the probabilities of the outcomes are $\Pr(M) = \tr (M\rho)$ and $\Pr(\id - M) = \tr((\id - M)\rho) = 1 - \Pr(M)$. The optimal bias of distinguishing two states $\rho$ and $\sigma$ is then given by $\max_{0 \leq M \leq \id}\tr(M(\rho - \sigma)) = \frac{1}{2}\Vert \rho - \sigma \Vert_{1}$.}  
In analogy with this interpretation of the trace norm, we define the LOCC norm as \cite{MWW09}
\begin{equation*}
\norm{X}_{\Mfont{LOCC}} := \max_{M \in \Mfont{LOCC}} \tr((2M - \id) X),
\end{equation*}
where $\LOCC$ is the convex set of matrices $0 \leq M \leq \id$ such that there is a two-outcome measurement $\{M, \id - M \}$ that can be realized by LOCC.\footnote{This defines a norm because the set of operators $2\Mclass- \id$ is convex, closed, symmetric about the origin and has nonempty interior.  Therefore it is the unit ball for a norm whose corresponding dual norm is equal to $\norm{*}_\LOCC$.} The optimal bias in distinguishing $\rho$ and $\sigma$ by any LOCC protocol is then $\frac{1}{2} \norm{\rho - \sigma }_{\Mfont{LOCC}}$. We note that in many applications of the separability problem, e.g.\ assessing the usefulness of a quantum state for violating Bell's inequalities or for performing quantum teleportation, the LOCC norm is actually the more relevant quantity to consider. 

The Euclidean, or Frobenius norm $\norm{X}_2 := \tr(X^{\cal y}X)$ is the negative exponential of the quantum collision entropy, and is often of interest in quantum information theory because its quadratic nature makes it especially easy to work with. 

The algorithm for testing separability, which we present and analyze in more detail in Section \ref{proofs}, is very simple and searches for symmetric extensions of the state using semidefinite programming. The search for symmetric extensions using semidefinite programming as a test of separability has first been proposed by Doherty, Parillo and Spedalieri \cite{DPS04}. 

\vspace{0.2 cm}
\noindent \textbf{A quasipolynomial-time algorithm for \textsc{Best Separable State}:} The same method used to prove Theorem \ref{main} also results in the following new algorithm for \textsc{BSS($\epsilon$)}:
\begin{theorem} \label{main2}
There is an algorithm solving \textsc{BSS($\epsilon$)} for the Hermitian operator $M$ in time 
$$\exp \left( O \left( \ep^{-2}\log|A| \log|B|  \Vert M \Vert_2^{2} \right)  \right).$$ Furthermore, there is an 
$\exp \left( O( \ep^{-2}\log|A| \log|B| )  \right)$-time algorithm solving \textsc{BSS($\epsilon$)} for any $M$ such that $\{ M, \id - M \}$ is an LOCC measurement.
\end{theorem}

It is intriguing that the complexity of our algorithm for LOCC operators $M$ matches the hardness result of Harrow and Montanaro for general operators, which shows that a subexponential-time algorithm of complexity up to 
\[\exp \left( O \left(\log^{1 - \nu} |A|   \log^{1 - \mu} |B| \Vert M \Vert_{\infty}\right) \right)\] for constant $\epsilon$ and any $\nu + \mu > 0$ would imply a $\exp \left(  o(n) \right)$ algorithm for $\SAT$ with $n$ clauses.\footnote{In fact, the operator $M$ can be taken to be a non-normalized separable state \cite{HM10}. This, however, does not imply that it can be implemented by LOCC.} It is an open question if a similar hardness result could be obtained for LOCC measurements, which would imply that our algorithm is optimal, assuming $\SAT$ requires exponential time.

\begin{sloppy}
An application of Theorem~\ref{main2} concerns the estimation of the ground state energy of mean-field Hamiltonians. A mean-field Hamiltonian consists of a Hermitian operator acting on $n$ sites (each formed by a $d$-dimensional quantum system) defined as $H := \frac{2}{n} \sum_{i < j} K_{i, j}$, with $K_{i, j}$ given by the Hermitian matrix which acts as $K$ on sites $i$ and $j$ (for a fixed two-sites interaction $K$) and as the identity on the remaining sites. Mean-field Hamiltonians are often used in condensed-matter physics as a substitute for a given local Hamiltonian, since they are easier to analyze and in many cases provide a good approximation to the true model. 
\end{sloppy}

An important property of quantum many-body Hamiltonians is their ground-state energy, i.e.\ their minimal eigenvalue. A folklore result in condensed-matter physics, formalized e.g.\ in \cite{FV06}, is that the computation of the ground-state energy of a mean-field Hamiltonian $H$ is equivalent to the minimization of $\tr(\sigma K)$ over separable states $\sigma \in \CS_{A:B}$ with $|A| = |B| = d$. Theorem \ref{main2} then readily implies a $\exp \left( O \left( \epsilon^{-2}  \log^{2}(d)  \Vert K \Vert_{2}^{2} \right)  \right)$-time algorithm for the problem. Before, the best-known algorithm\footnote{The algorithm again simply searches for the minimum overlap of $K$ over an $\epsilon$-net in the set of product states.} scaled as  
$$\exp \left( \Omega \left( d \Vert K \Vert_{\infty} \log(\epsilon^{-1}) \right)  \right).$$

\vspace{0.2 cm}
\noindent \textbf{Monogamy of entanglement and LOCC norm:} We say that a bipartite state $\rho_{A:B}$ is $k$-extendible if there is a state $\rho_{A:B_1\cdots B_k}$ that is permutation-symmetric in the $B$ systems with $\rho_{A:B} = \tr_{B_2\cdots B_k}(\rho_{A:B_1\cdots B_k}) $. The sets of $k$-extendible states provide a sequence of approximations to the set of separable states. In the limit of large $k$, the approximation becomes exact because a state is separable if, and only if, it is $k$-extendible for every $k$ (see e.g.\ \cite{CKMR07}). This result is a manifestation of a property of quantum correlations known as \textit{monogamy of entanglement}: a quantum system cannot be equally entangled with an arbitrary number of other systems, i.e.\ entanglement is a non-shareable property of quantum states. 

In a quantitative manner, quantum versions of the de Finetti theorem imply that for any $k$-extendible state $\rho_{A:B}$:  $\norm{\rho_{A:B} - \CS}_1 \leq 4|B|^{2}k^{-1}$.\footnote{The quantum de Finetti theorem in \cite{CKMR07} says that given a $(k+1)$-partite quantum state $\rho_{AB_1, \ldots ,B_k}$ invariant under exchange of the systems $B_i$, there is a measure $\mu$ on quantum states on system $B$ such that $\left \Vert \rho_{AB_1, \ldots, B_l}  - \int \mu(d\sigma) \xi^\sigma_A \otimes \sigma_{B_1} \otimes \cdots \otimes \sigma_{B_l} \right \Vert_1 \leq 4d^{2}l/k$.} Moreover, this bound is close to tight, as there are $k$-extendible states that are $\Omega(|B|k^{-1})$-away from the set of separable states \cite{CKMR07}. Unfortunately, for many applications this error estimate -- exponentially large in the number of qubits of the state -- is too big to be useful. The key result behind Theorems \ref{main} and \ref{main2} is the following de Finetti-type result, which shows that a significant improvement is possible if we are willing to relax our notion of distance of two quantum states:
\begin{theorem} \label{monogamy}
Let $\rho_{A:B}$ be $k$-extendible. Then 
\[\norm{\rho_{A:B} - \CS}_\leftrightLOCC \leq  O\!\left(k^{-1}\log |A| \right)^{\frac{1}{2}}.\]
\end{theorem}
\noindent
In \cite{MWW09} it was shown that $\Vert X \Vert_{\leftrightLOCC} \geq \frac{1}{\sqrt{153}} \Vert X \Vert_2$, so we also have a similar bound for the Euclidean norm, namely
\begin{equation} \label{boundeuclidean}
\norm{\rho_{A:B} - \CS}_2 \leq  O\!\left(k^{-1}\log |A| \right)^{\frac{1}{2}}
\end{equation}

A direct implication of Theorem \ref{monogamy} concerns data-hiding states \cite{DLT02, DHT03, EW02, HLW06}. Every state $\rho$ that can be well-distinguished from separable states by a global measurement, yet is almost completely indistinguishable from a separable state by LOCC measurements is a so-called data-hiding state: it can be used to hide a bit of information (whether the prepared state is $\rho$ or the closest separable state to $\rho$ in LOCC norm) that is not accessible by LOCC operations alone. The bipartite antisymmetric state of sufficiently high dimension is an example of a data hiding state \cite{EW02}, as are random mixed states with high probability \cite{HLW06} (given an appropriate choice of the dimensions and the rank of the state). Theorem \ref{monogamy} shows that highly extendible states that are far away in trace norm from the set of separable states must necessarily be data-hiding.

\vspace{0.2 cm}
\noindent \textbf{Quantum Merlin-Arthur games with multiple Merlins:} A final application of Theorem \ref{monogamy} concerns the complexity class Quantum Merlin-Arthur ($\QMA$), the quantum analogue of $\NP$ (or more precisely of $\MA$). It is natural to ask how robust the definition  of $\QMA$ is and a few results are known in this direction: For example, it is possible to amplify the soundness and completeness parameters to exponential accuracy, even without enlarging the proof size \cite{MW05}. Also, the class does not change if we allow a first round of logarithmic-sized quantum communication from the verifier to the prover \cite{BSW10}. 

From Theorem \ref{main2} we get a new characterization of $\QMA$, which at first sight might appear to be strictly more powerful: We show $\QMA$ to be equal to the class of languages that can be decided in polynomial time by a verifier who is given $k$ unentangled proofs and can measure them using any quantum polynomial-time implementable LOCC protocol among the $k$ proofs. This answers an open question of Aaronson \textit{et al.} \cite{ABDFS08}. We hope this characterization of $\QMA$ proves useful in devising new $\QMA$ verifying systems.

\begin{sloppy}
In order to formalize our result, let $\Mclass$ be a class of two-outcome measurements and consider the classes $\QMA_{\Mclass}(k)_{m,s,c}$, defined in analogy to $\QMA$ as follows \cite{KMY03, ABDFS08, HM10}:
\begin{definition} \label{defQMA2}
A language $L$ is in $\QMA_{\Mclass}(k)_{m, s, c}$ if 
there is a uniform family of polynomial-sized quantum circuits that,
for every input $x \in \{ 0, 1\}^{n}$, can implement a two outcome measurement $\{ M_x, \id - M_x \}$ from the class $\Mclass$ such that
\begin{itemize}
\item \emph{Completeness:} If $x \in L$, there exist $k$ witnesses $\ket{\psi_1},\dotsc,\ket{\psi_k}$, each of $m$ qubits, such that
\begin{equation*}
\tr \left( M_x \left(\ket{\psi_1}\bra{\psi_1} \otimes \cdots \otimes \ket{\psi_k}\bra{\psi_k} \right)  \right) \geq c.
\end{equation*}
\item \emph{Soundness:} If $x \notin L$, then for any states $\ket{\psi_1}, \dotsc, \ket{\psi_k}$
\begin{equation*}
\tr \left( M_x \left(\ket{\psi_1}\bra{\psi_1} \otimes \cdots \otimes \ket{\psi_k}\bra{\psi_k} \right)  \right) \leq s.
\end{equation*}
\end{itemize}
We call $\QMA_{\Mclass}(k) = \QMA_{\Mclass}(k)_{\poly(n), 2/3, 1/3}$.
By a \emph{uniform family}, we mean that there should be a classical algorithm which, upon given the input length $n$ and the string $x$, outputs a description of the quantum circuit implementing the measurement $\{M_x,\id - M_x\}$ in time $O(\poly(n))$.
\end{definition}
\end{sloppy}
Let $\SEP_{\text{YES}}$ be the class of two outcome POVMs $\{M,\id - M\}$ such that $M$, the POVM element corresponding to \emph{accept}, is a (non-normalized) separable operator.
Harrow and Montanaro showed that 
\[\QMA_{\SEP_{\text{YES}}}(2) = \QMA(2) = \QMA(k)\] for any $k = \poly(n)$ \cite{HM10}, i.e.\ two proofs are just as powerful as $k$ proofs and one can  restrict the verifier's action to $\SEP_{\text{YES}}$ without changing the expressive power of the class. 

We define $\QMA_{\leftrightLOCC}(k)$ in an analogous way, but now the verifier can only measure the $k$ proofs with a LOCC measurement. Then we have,
\begin{theorem} \label{QMA2}
For $k = O(1)$,
\begin{equation} \label{QMA2LOCCequalQMA}
\QMA_{\leftrightLOCC}(k) = \QMA .
\end{equation}
In particular,
\begin{equation} \label{qma2exact}
\QMA_{\leftrightLOCC}(2)_{m, s, c} \subseteq \QMA_{ O(m^{2} \epsilon^{-2}), s + \epsilon, c} .
\end{equation}
\end{theorem}

A preliminary step in the direction of Theorem \ref{QMA2} appeared in \cite{Bra08}, where a similar result was shown for $\QMA_{\LO}(k)$, a variant of $\QMA(k)$ in which the verifier is restricted to implement only local measurements on the $k$ proofs and jointly post-process the outcomes classically.\footnote{$\QMA_{\LO}(k)$ is also called $\BellQMA(k)$ \cite{ABDFS08} since the verifier is basically restricted to perform a \textit{Bell test} on the proofs.}

It is an open question whether Eq.~(\ref{qma2exact}) remains true if we consider $\QMA(2)$ instead of $\QMA_{\leftrightLOCC}(2)$. If this turns out to be the case, then it would imply an optimal conversion of $\QMA(2)$ into $\QMA$ in what concerns the proof length (under a plausible complexity-theoretic assumption). For it follows from \cite{HM10} (based on the $\QMA(\sqrt{n}\polylog (n))_{\log(n), 1/3, 1}$ protocol for 3-$\SAT$ with $n$ variables of \cite{ABDFS08}) that unless there is a subexponential-time quantum algorithm for 3-$\SAT$, then there is a constant $\epsilon_0 > 0$ such that for every $\delta > 0$, 
\[\QMA(2)_{m, s, c} \nsubseteq \QMA_{ O(m^{2-\delta}), s + \epsilon_{0}, c}.\] 

Recently Chen and Drucker \cite{CD10} showed that a variant of the 3-$\SAT$ protocol from \cite{ABDFS08} can be implemented with only local measurements, showing\footnote{In fact they proved the stronger statement that 3-$\SAT$ is in $\QMA_{\LO}(\sqrt{n}\polylog(n))_{\log(n), 1/3, 2/3}$.} that 3-$\SAT$ is in 
\[\QMA_{\leftrightLOCC}(\sqrt{n}\polylog(n))_{\log(n), 1/3, 2/3}.\] It is an intriguing open question if one could also obtain a $\QMA_{\leftrightLOCC}(2)$ protocol with the same total proof length ($O(\sqrt n \polylog(n))$), which would imply that the reduction from $\QMA_{\leftrightLOCC}(2)$ to $\QMA$ given in Theorem \ref{QMA2} cannot be improved, unless there is a subexponential time quantum algorithm for $\SAT$.

\vspace{0.2 cm}
\begin{sloppy}
We will now give a characterization of $\QMA$ in terms of protocols for multiple provers with a restriction on the Euclidean norm of the verifiers measurements. Let $\QMA_{\sf{LOW}, no}(k)$ be defined as above, with $\sf{LOW}$ the class of measurements $\{ M, \id - M \} $ for which $\Vert M \Vert_2 \leq \poly(n)$, but with such a restriction imposed only on the \textit{no} instances of the language. 
\begin{definition}
A language $L$ belongs to $\QMA_{\sf{LOW}, no}(k)$ if there is a uniform family of quantum circuits that, for every $x \in \{0,1\}^n$, can implement a two-outcome measurement $\{ M_x, \id - M_x \}$ 
such that
\begin{itemize}
\item \emph{Completeness:} If $x \in L$, there exist $k$ witnesses $\ket{\psi_1},\dotsc,\ket{\psi_k}$, each of $\poly(n)$ qubits, such that
\begin{equation*}
\tr \left( M_x \left(\ket{\psi_1}\bra{\psi_1} \otimes \cdots \otimes \ket{\psi_k}\bra{\psi_k} \right)  \right) \geq \frac{2}{3}.
\end{equation*}
\item \emph{Soundness:} If $x \notin L$, then $\Vert M_x \Vert_2 \leq \poly(n)$ and for any states $\ket{\psi_1}, \dotsc, \ket{\psi_k}$
\begin{equation*}
\tr \left( M_x \left(\ket{\psi_1}\bra{\psi_1} \otimes \cdots \otimes \ket{\psi_k}\bra{\psi_k} \right)  \right) \leq \frac{1}{3}.
\end{equation*}
\end{itemize}
\end{definition}
Then we also have 
\begin{theorem} \label{lowQMA}
For $k = O(1)$,
\begin{equation*}
\QMA_{\sf{LOW}, no}(k)= \QMA. 
\end{equation*}
\end{theorem}

\end{sloppy}

It is an open question whether Theorems~\ref{QMA2} and \ref{lowQMA} hold for nonconstant $k$, say for $k = O(\poly(n))$.  Our methods fail to achieve this because the quadratic blowup in the proof size inherent to our proofs prevents us from applying the reduction recursively more than a constant number of times.

\vspace{0.2 cm}
\noindent \textbf{Existence of disentanglers}: An interesting approach to the $\QMA(2)$ vs.\ $\QMA$ question concerns the existence of disentangler superoperators \cite{ABDFS08}, defined as follows: A superoperator $\Lambda : S \rightarrow A \otimes B$ is an $(\log|S|, \epsilon, \delta)$-disentangler in the $\Vert * \Vert$ norm if 
\begin{itemize}
\item  $\Lambda(\rho)$ is $\epsilon$-close to a separable state for every $\rho$, and
\item  for every separable state $\sigma$, there is a $\rho$ such that $\Lambda(\rho)$ is $\delta$-close to $\sigma$.
\end{itemize}
As noted in \cite{ABDFS08}, the existence of an efficiently implementable $(\poly(\log |A|, \log |B|), \epsilon, \delta)$-disentangler in trace norm (for sufficiently small $\epsilon$ and $\delta$) would imply $\QMA(2) = \QMA$. Watrous has conjectured that this is not the case and that for every $\epsilon, \delta < 1$, any $(\epsilon, \delta)$-disentangler (in trace-norm) requires $|S| = 2^{\Omega( \min(|A|, |B|))}$. 

Theorem \ref{monogamy} readily implies that the LOCC-norm analog of Watrous' conjecture fails: there \textit{is} an efficient disentangler in LOCC norm. Indeed, let $k = \Omega \left( \ep^{-2}\log |A|  \right)$ and $S := a \otimes b_1 \otimes \cdots \otimes b_k$. Define the superoperator $\Lambda : S \rightarrow A \otimes B$, with $|A| = |a|$ and $|B| = |b_j|$ for all $j \leq k$, as
\begin{equation*}
\Lambda(\rho_{a b_1\cdots b_k}) := \frac{1}{k} \sum_{i=1}^{k} \rho_{a b_i}.
\end{equation*}
Then $\Lambda$ is a $(O( \epsilon^{-2}\log |A| \log |B|), \epsilon, 0)$-disentangler in LOCC norm.  

\vspace{0.2 cm}
\noindent \textbf{A lower bound on conditional mutual information:} The main technical tool we use for obtaining Theorem \ref{monogamy} is a new lower bound on the quantum conditional mutual information of tripartite quantum states $\rho_{ABE}$, which might be of independent interest. The conditional mutual information is defined as 
\[I(A;B|E)_{\rho} := H(AE)_{\rho} + H(BE)_{\rho} - H(ABE)_{\rho} - H(E)_{\rho},\] where $H(X)_{\rho} := -  \tr(\rho_X \log \rho_X)$ is the von Neumann entropy. Then we have the following analog of Pinsker's inequality\footnote{Pinsker's inequality for the relative entropy implies that $I(A;B) \geq \frac 1{2\ln 2}\norm{\rho_{AB} - \rho_A\ox \rho_B}_1^{2}.$}:
\begin{theorem}  \label{boundCMI}
For every $\rho_{ABE}$, 
\[I(A;B|E)_{\rho} \geq \frac{1}{8 \ln 2}\norm{\rho_{A:B} - \CS}_\LOCC^2.\]
\end{theorem}

Theorem \ref{boundCMI} leads to a new result concerning the entanglement measure \textit{squashed entanglement} \cite{CW04}, defined as 
\begin{equation*}
E_{sq}(\rho_{A:B}) = \inf \big\{ \smfrac{1}{2} I(A;B|E) : \hspace{0.1 cm} \rho_{ABE} \hspace{0.1 cm} \text{extension of} \hspace{0.1 cm} \rho_{AB}  \big\}.
\end{equation*}
An immediate corollary of Theorem~\ref{boundCMI} is then 
\begin{corollary} \label{squashed}
For every $\rho_{AB}$, 
\[E_{sq}(\rho_{A:B})  \geq \frac{1}{16 \ln 2}\norm{\rho_{A:B} -\CS}_{\leftrightLOCC}^2.\]
\end{corollary}
In particular, this implies that squashed entanglement is \emph{faithful}, meaning it is strictly positive on all entangled states. This had been a long-standing conjecture in entanglement theory.

\section{Proofs} \label{proofs}
We now give complete proofs of our theorems with the exception of Theorem~\ref{boundCMI}, for which we give a brief outline of the proof strategy. A complete proof can be found in~\cite{BCY10}. We begin with a brief proof of Theorem \ref{monogamy}, which itself is the key for the complexity-theoretic results. 
\begin{proof}[of Theorem \ref{monogamy}]
This theorem is a simple combination of Corollary \ref{squashed} and the following monogamy relation for squashed entanglement: For every bipartite state $\rho_{A:B_1\cdots B_k}$ \cite{KW04}:
\begin{equation*} 
\log(|A|) \geq E_{sq}(\rho_{A:B_1\cdots B_k}) \geq \sum_{j=1}^{k} E_{sq}(\rho_{A:B_j}).
\end{equation*}
This and Corollary \ref{squashed} give the proof.
\end{proof}

\begin{proof}[of Theorem \ref{main}]

 We prove the statement for the LOCC norm. The Euclidean norm case follows by the same argument, replacing each application of Theorem \ref{monogamy} by Eq. (\ref{boundeuclidean}).

The idea of the algorithm, which is also the basic idea of the algorithm from \cite{DPS04}, is to formulate the search for a $O\left( \ep^{-2}\log|A|\right)$-extension of $\rho_{AB}$ as a semidefinite program (SDP) \cite{VB96}. If $\rho_{AB}$ is separable then such an extension exists because separable states have a $k$-extension for every $k$. Otherwise if $\norm{\rho_{AB} - \CS}_\LOCC \geq \epsilon$, no such extension exists by Theorem \ref{monogamy}. We only have to make sure that the precision of the algorithm solving the SDP is good enough, which we now analyze in detail. 

Consider the following semidefinite program, with $\tau_{A:B} = \id / (|A| |B|)$ the maximally mixed state, $\delta := \epsilon/2$ and $\rho_{AB, \delta} :=  (1 - \delta)\rho_{A:B} + \delta \tau_{A:B}$,
\begin{equation}
 \label{SDPproblem}
\max_{X} \tr(X_{A:B_1\cdots B_k}) 
\text{ subject to:} \begin{array}{c} X_{A:B_1\cdots B_k} \geq 0 \\ X_{A:B_j} \leq \rho_{AB, \delta} \hspace{0.2 cm} \forall j. \end{array}
\end{equation}
We introduced $\rho_{AB, \delta}$ as we require a non-negligible bound on the minimum eigenvalue of the state.  Observe that $\rho_{AB,\delta}$ has a $k$-extension precisely when the solution of (\ref{SDPproblem}) is 1, in which case the extension is obtained by symmetrizing the $B$ parts of $X$, i.e.\ by replacing $X$ with the operator 
\[\frac{1}{k!}\sum_{\pi \in S_k} (\id_A \ox \pi_B) X(\id_A \ox \pi_B)^{-1},\]
where the sum is over permutations. 
 
We now consider the approximate case.  Define 
\[{\cal F} := \{ X_{A:B_1\cdots B_k} : X \geq 0, X_{A:B_j} \leq \rho_{A:B, \delta} \hspace{0.2 cm} \forall j \in [k] \}\] as the set of feasible points and ${\cal F}_{\nu}$ its $\nu$-interior, i.e.\ 
\[{\cal F}_{\nu} := \{ X_{A:B_1\cdots B_k} : X + H \in {\cal F} \hspace{0.2 cm} \text{for all} \hspace{0.1 cm} H \hspace{0.2 cm} \text{s.t.} \hspace{0.2 cm} \Vert H \Vert_2 \leq \nu \}.\] 
The use of Frobenius norm in the definition of ${\cal F}_\nu$ is completely independent of the norm in the theorem statement.  Rather, it ensures  
the ellipsoid algorithm solves problem (\ref{SDPproblem}) up to additive error $\nu$ in time $\poly(|A||B|^{k}, \log(\nu^{-1}))$ as long as ${\cal F}_{\nu}$ is nonempty (see e.g.~\cite{Wat09} and references therein).  We claim that $\mathcal{F}_\nu$ is nonempty when $\nu := \exp(-|A||B| \epsilon^{-2})$ and $k=O(\epsilon^{-1} \log |A|)$.  Before proving this, let us show how it implies that we can solve the weak-membership problem for separability by solving (\ref{SDPproblem}).

Suppose first that $\rho_{A:B}$ is separable.  Convexity of $\CS$ implies that $\rho_{A:B, \delta}$ is also separable, so we know there is a symmetric extension $\rho_{A:B_1\cdots B_k, \delta}$ of $\rho_{A:B, \delta}$. The ellipsoid algorithm applied to problem (\ref{SDPproblem}) will therefore return a number bigger than $1 - \nu$. 

Suppose now that $\rho_{A:B}$ is $\epsilon$-away from $\CS$. Then $\rho_{A:B, \delta}$ is $\epsilon/2$-away from $\CS$. By Theorem~\ref{monogamy}, any state $\tilde \rho_{A:B}$ that is $\epsilon/4$-close to $\rho_{A:B, \delta}$ in LOCC norm does not have a $O\left( \ep^{-2}\log|A|\right)$-extension. From this we can get that the solution of the SDP (\ref{SDPproblem}) will be smaller than $1 - \Omega(\epsilon)$. Indeed suppose it were not the case and that the solution was larger than $1 - c\epsilon$ (for sufficiently small $c > 0$). Then because we are guaranteed to be at most $\nu$ away from the exact solution of (\ref{SDPproblem}), this would imply there is a positive semidefinite matrix $Y_{A:B_1\cdots B_k}$ such that $Y_{A:B_j} \leq \rho_{A:B, \delta}$ for every $j \in [k]$ and $\tr(Y) \geq 1 - (c\epsilon + \nu)$. We can symmetrize the $B$ systems in $Y_{A:B_1\cdots B_k}$ to obtain a semidefinite positive matrix $Z_{A:B_1\cdots B_k}$, symmetric under the exchange of the $B$ systems and such that $Z_{A:B_1} \leq \rho_{A:B, \delta}$  and $\tr(Z) \geq 1 - (c\epsilon + \nu)$. Defining $\sigma_{A:B_1\cdots B_k} = Z/\tr(Z)$, we find $\sigma_{A:B_1}$ to be $k$-extendible with $\sigma_{A:B_1} \leq (1 + 2(c\epsilon + \nu))\rho_{A:B, \delta}$, so $\Vert \sigma_{A:B_1} - \rho_{A:B, \delta} \Vert_1 \leq 4(c\epsilon + \nu)$. But this is a contradiction, since we found before that the $\epsilon/4$-ball around $\rho_{A:B, \delta}$ does not contain any $k$-extendible state. Because $k = O(\ep^{-2}\log|A|)$, the computational cost of solving the ellipsoid algorithm with accuracy $\nu = \exp(-|A||B|\epsilon^{-2})$ is 
\begin{eqnarray*}
\polylog(1/\nu) \poly(|A||B|^{k}) \hspace{-1.2in} \\ 
&=& \exp \left( O(\epsilon^{-2} \log|A| \log |B|) + O(\log (|A||B|\epsilon^{-1})) \right) \\
&=& \exp \left( O(\epsilon ^{-2} \log |A| \log|B|)) \right).
\end{eqnarray*} 
\begin{sloppy}
We now prove that ${\cal F}_{\nu}$ is nonempty. This follows from the fact that $T_{A:B_1\cdots B_k} := \frac{\epsilon}{4}\tau_{A:B_1\cdots B_k} \in {\cal F}_{\nu}$, where $\tau_{A:B_1\cdots B_k} := \id/ |A| |B|^{k}$ the maximally mixed state. Indeed, it is clear that $T_{A:B_1\cdots B_k} + H \geq 0$ for every 
$$\Vert H \Vert_\infty \leq \Vert H \Vert_2  \leq \exp(-|A||B|\epsilon^{-2}).$$ Moreover, $T_{A:B_1\cdots B_k} - H \geq 0$ which immediately implies that $\tr_{B_2\cdots B_k}(T + H) \leq \tr_{B_2\cdots B_k}(2T)\leq \rho_{A:B, \delta}$.
\end{sloppy}
\end{proof}

\begin{proof}[of Theorem \ref{main2}]

Let ${\cal E}_k$ be the set of $k$-extendible states. Let us first analyze the case in which $M$ is such that $\{M, \id - M \}$ is LOCC. Then the inclusion ${\cal S} \subset {\cal E}_k$ and Theorem \ref{monogamy} give
\begin{eqnarray*}
\max_{\rho \in {\cal E}_k} \tr(\rho M) &\geq& \max_{\sigma \in {\cal S}} \tr(\sigma M) \nn \\
&\geq& \max_{\rho \in {\cal E}_k} \tr(\rho M) - \sqrt{O\left(k^{-1} \log |A|  \right)}. 
\end{eqnarray*}
Hence choosing $k = O(\epsilon^{-2} \log|A|)$ we can compute an $\epsilon$-error additive approximation to \textsc{BSS($\epsilon$)} by solving the semidefinite program given by maximizing $\tr(M \rho)$ over $k$-extendible states, whose time-complexity is $\exp \left( O \left( \epsilon^{-2}\log |A| \log|B|   \right)  \right)$. This proves the first part of the theorem. 

To obtain the bound for general $M$, note that 
\[|\!\tr(M(\rho - \sigma))| \leq \Vert M \Vert_2 \Vert \rho - \sigma \Vert_2\] by the Cauchy-Schwarz inequality. Therefore
\begin{eqnarray*}
\max_{\rho \in {\cal E}_k} \tr(\rho M) &\geq& \max_{\sigma \in {\cal S}} \tr(\sigma M) \nn \\ &\geq& \max_{\rho \in {\cal E}_k} \tr(\rho M) - \Vert M \Vert_2\,  \sqrt{O\!\left(k^{-1} \log |A|  \right)}. \hspace{.25in}
\end{eqnarray*}
Then choosing $k = O(\epsilon^{-2} \log|A| \Vert M \Vert_2^{2})$ we can obtain an $\epsilon$-error additive approximation to \textsc{BSS($\epsilon$)} by solving a SDP of time-complexity $\exp \left( O \left(\epsilon^{-2} \Vert M \Vert_2^{2}  \log |A| \log|B|  \right)  \right)$
\end{proof}


\begin{proof}[of Theorem \ref{QMA2}]

We start by proving Eq.~(\ref{qma2exact}). Consider a protocol in $\QMA_{\leftrightLOCC}(2)_{m, s, c}$ given by the LOCC measurement $\{M, \id - M\}$. We construct a $\QMA_{O(m^{2}\epsilon^{-2}), s + \epsilon, c}$ protocol that can simulate it: The verifier asks for a proof of the form $\ket{\psi}_{A:B_1\cdots B_k}$ where $|A| = |B_j| = 2^{m}$ (each register consists of $m$ qubits) and $k = \Omega(m\epsilon^{-2})$. He then symmetrizes the $B$ systems obtaining the state $\rho_{A:B_1\cdots B_k}$ and measures $\{M, \id - M\}$ in the subsystems $AB_1$. 

Let us analyze the completeness and soundness of the protocol. For completeness, the prover can send $\ket{\psi}_{A} \otimes \ket{\phi}_B^{\otimes k}$, for states $\ket{\psi}, \ket{\phi}$ such that $\tr \proj{\phi} \otimes \proj{\psi} M \geq c$. Thus the completeness parameter of the $\QMA$ protocol is at least $c$. 

For soundness, we note that $\norm{\rho_{A:B_1} - \CS}_{\leftrightLOCC} \leq \epsilon$ by Theorem  \ref{monogamy}. Thus, as $\{ M, \id - M\} \subset \leftrightLOCC$ the soundness parameter for the $\QMA$ protocol can only be $\epsilon$ away from $s$. Indeed, for every $\rho_{A:B_1\cdots B_k}$ symmetric in the $B$ systems, 
\[\tr(\rho_{A:B_1} M) \leq \max_{\sigma \in {\cal S}} \tr(M \sigma) +  \norm{\rho_{A:B_1} - \CS}_{\leftrightLOCC} \leq s + \epsilon.\]

Eq.~(\ref{QMA2LOCCequalQMA}) follows from the protocol above. Given a protocol in $\QMA_{\leftrightLOCC}(\ell)$ with each proof of size $m$ qubits we can simulate it in $\QMA_{\leftrightLOCC}(\ell-1)$ as follows: The verifier asks for $\ell-1$ proofs, the first proof consisting of registers $AB_1\cdots B_k$, each of size  $m$ qubits and $k = \Omega(m \epsilon^{-2})$, and all the $\ell-2$ other proofs of size $m$ qubits. Then he symmetrizes the $B$ systems and traces out all of them except the first. Finally he applies the original measurement from the $\QMA_{\leftrightLOCC}(\ell)$ to the resulting state.

The completeness of the protocol is unaffected by the simulation. For the soundness let $\rho_{AB_1\cdots B_k} \otimes \sigma_3 \otimes \cdots \otimes \sigma_{l}$ be an arbitrary state sent by the prover (after symmetrizing $B_1,\dotsc, B_m$). Let $\{ M, \id - M  \} \subset \leftrightLOCC$ be the verification measurement from the $\QMA_{\leftrightLOCC}(l)$ protocol. Then
\begin{eqnarray*}
\tr \left(A (\rho_{AB_1} \otimes \sigma_3 \otimes \cdots \otimes \sigma_{l}) \right) \hspace{-1.5in} \\
&\leq& \max_{\sigma \in {\cal S}(1:2:\cdots :l)} \tr(M \sigma) \\
&& \qquad + \min_{\sigma \in {\cal S}} \Vert \rho_{AB_1} \otimes \sigma_3 \otimes \cdots \otimes  \sigma_{l} - \sigma \Vert_{\leftrightLOCC} \nonumber \\ &=& \max_{\sigma \in {\cal S}(1:2:\cdots :l)} \tr(M \sigma) + \min_{\sigma \in {\cal S}} \Vert \rho_{AB_1} - \sigma \Vert_{\leftrightLOCC} \nonumber \\ &\leq& s + \epsilon,
\end{eqnarray*}
The equality in the second line follows since we can assume that the states $\sigma_{3}, \dotsc , \sigma_{l}$ belong to the verifier and adding local states does not change the minimum LOCC-distance to separable states. 

Since for going from $\QMA_{\leftrightLOCC}(\ell)$ to $\QMA_{\leftrightLOCC}(\ell-1)$ we had to blow up one of the proof's size only by a quadratic factor, we can repeat the same protocol a constant number of times and still get each proof of polynomial size. In the end, the completeness parameter of the $\QMA$ procedure is the same as the original one for $\QMA_{\leftrightLOCC}(\ell)$, while the soundness is smaller than $s + \ell \epsilon$, which can be taken to be a constant away from $c$ by choosing $\epsilon$ sufficiently small. To reduce the soundness back to the original value $s$ we then use the standard amplification procedure for $\QMA$ (see e.g.\ \cite{Wat08}), which works in this case since the verification measurement is LOCC \cite{ABDFS08}. 
\end{proof}

\begin{proof}[of Theorem \ref{lowQMA}]
The proof is very similar to the proof of Theorem \ref{QMA2}, so we only comment on the differences. The strategy for simulating a $\QMA_{\sf{LOW}, no}(2)$ protocol in $\QMA$ is the same as before: The verifier asks for a proof of the form $\ket{\psi}_{A:B_1\cdots B_k}$ where $|A| = |B_j| = 2^{m}$ (each register consists of $m$ qubits) and $k = \poly(n)\epsilon^{-2}$. He then symmetrizes the $B$ systems to obtain the state $\rho_{A:B_1\cdots  B_k}$, and measures $\{M, \id - M\}$ in the subsystems $AB_1$. The completeness of the $\QMA$ protocol is the same as that of the original, since the prover can send $\ket{\psi}_{A} \otimes \ket{\phi}_B^{\otimes k}$.

For analyzing the soundness of the protocol, let $\sigma$ be the closest separable state to $\rho_{A:B_1}$ in Euclidean norm. Eq.~(\ref{boundeuclidean}) gives 
\[\Vert \rho_{A:B_1} - \sigma \Vert_2 \leq O\!\left(k^{-1}\log |A| \right)^{\frac{1}{2}} = 1/\poly(n).\] Then, by the Cauchy-Schwarz inequality, 
\[|\tr((\rho_{A:B_1} - \sigma)M_x)| \leq  \Vert \rho_{A:B_1} - \sigma \Vert_2 \Vert M_x \Vert_2 \leq 1/\poly(n).\] The proof for $\QMA_{\sf{LOW}, no}(k)$ for $k > 2$ is completely analogous to the proof of Theorem \ref{QMA2}.

A last point to argue is the converse relation, namely that $\QMA$ is contained in $\QMA_{\sf{LOW}, no}(2)$. This follows from the $\QMA$ error reduction protocol of Marriot and Watrous \cite{MW05}. Indeed, they showed how any protocol in $\QMA$ can be transformed into a protocol with proof size $n$ equal to the original proof size and soundness $2^{-\poly(n)}$. This means that for ``no" instances the associated measurement $M_x$ must be such that $\Vert M_x \Vert_2 \leq 2^{n} \Vert M_x \Vert_{\infty} \leq 2^{-\poly(n)}$, from which follows that the protocol is in $\QMA_{\sf{LOW}, no}(2)$.
\end{proof}

\begin{proof}[of Theorem \ref{boundCMI} (outline)]   The proof of Theorem \ref{boundCMI} begins by first chaining together three inequalities (Lemmas~\ref{nonlockability}, \ref{almostmonogamy} and \ref{lowerboundnorm} below), each of which is a new result in entanglement theory and is of independent interest.  A recursive step (Lemma~\ref{recursion} below) completes the proof.  These same lemmas appear in \cite{BCY10} with complete proofs; here we only outline these proofs. 

The first step involves an entanglement measure called the \emph{regularized relative entropy of entanglement} \cite{VPRK97}, defined as 
\begin{equation*}
E_{R}^{\infty}(\rho_{A:B}) \equiv \lim_{n \rightarrow \infty} \frac 1n E_{R}(\rho_{A:B}^{\otimes n}),
\end{equation*}
where $E_R(\rho_{A:B}) \equiv \min_{\sigma \in {\cal S}} S(\rho || \sigma)$ is the \emph{relative entropy of entanglement}, and where $S(\rho || \sigma) = \tr(\rho(\log \rho - \log \sigma))$ is the quantum \emph{relative entropy}.  

A distinctive property of the relative entropy of entanglement among entanglement measures is the fact that it is not ``lockable,'' meaning that after discarding a small part of the state, $E_R$ can only drop by an amount proportional to the number of qubits traced out. Indeed, as shown in \cite{HHHO05}, 
\begin{equation}
E_R(\rho_{A:BE}) \leq  E_R(\rho_{A:E}) +   2 S(B)_{\rho}. \label{nonlockable}
\end{equation}
While the same is true for $E_R^{\infty}$, we prove the following stronger version:
\begin{lemma} For every $\rho_{ABE}$,
\[I(A;B|E) \geq E_R^\infty(\rho_{A:BE}) -  E_R^\infty(\rho_{A:E}).\]
\label{nonlockability}
\end{lemma} 
\begin{proof}[(outline)]
This lemma follows by combining the inequality (\ref{nonlockable}) with an optimal protocol  for the following multipartite quantum data compression problem \cite{YD07}.  Consider many copies of a pure state\footnote{A \emph{pure state} is a rank 1 density matrix.  A basic theorem in quantum information asserts that every density matrix $\rho_A$ can be expressed as the partial trace of a pure state on a larger system $AB$.} on $ABEE'$ whose restriction to $ABE$ is $\rho_{ABE}$.  Suppose these states are shared between two parties: a sender, who holds $BE$, and a receiver, who holds $E'$, while $A$ is inaccessible to both.  The \emph{state redistribution problem} asks the sender to use quantum communication\footnote{The sender and receiver are also allowed to utilize shared entanglement between themselves to accomplish this task.} to transfer the $B$ system to the receiver, while asymptotically preserving the overall global quantum state.  A protocol for state redistribution was given in \cite{YD07} achieving the optimal communication rate of $\frac 12 I(A;B|E)$, providing an operational interpretation for quantum conditional mutual information.  The proof of Lemma~\ref{nonlockability} is obtained by carefully using the state redistribution protocol to apply the inequality (\ref{nonlockable}) to a tensor-power state $\rho^{\ox n}$ in the most efficient way.
\end{proof}

Next, we recall a recent operational interpretation of $E_R^{\infty}$ in the context of quantum hypothesis testing \cite{BP10}. Suppose Alice and Bob are given either $n$ copies of an entangled state $\rho_{A:B}$, or an arbitrary separable state across $A^{n} \!:\! B^{n}$.  Then we define $D_\Mclass(\rho_{AB})$ to be the optimal error exponent for distinguishing between these two situations, using only measurements from the class $\Mclass$.  Specifically, let 
\[p_e(n) = \min_M \max_{\sigma \in \CS_{A^n:B^n}} \tr (M \sigma),\]
where the minimization is over all  
measurements $M \in \Mclass$ identifying $\rho_{AB}^{\ox n}$ with asymptotically unit probability:  
\[\lim_{n \rightarrow \infty}\tr (M \rho_{AB}^{\ox n}) = 1.\]
\begin{equation*}
D_{\Mclass}(\rho_{A:B}) := \lim_{n \rightarrow \infty} - \frac{\log p_e(n)}{n}.
\end{equation*}
The main result of \cite{BP10} gives the following equality
\begin{equation} \label{DallequalsER}
D_{\ALL}(\rho_{A:B}) = E_{R}^{\infty}(\rho_{A:B}),
\end{equation}
i.e.\ the regularized relative entropy of entanglement is the optimal distinguishability rate when trying to distinguish many copies of an entangled state from (arbitrary) separable states, in the case where there is no restrictions on the measurements available. 

Define $\leftLOCC$ in analogy to $\LOCC$, using only measurements that can be implemented by \textit{one-way} LOCC, i.e.\ by any protocol formed by local operations and 
classical communication \textit{only} from Bob to Alice. Then we have:
\begin{lemma} For every $\rho_{ABE}$, 
\[E_R^\infty(\rho_{A:BE}) -  E_R^\infty(\rho_{A:E}) \geq D_{\leftLOCC}(\rho_{A:B}).\]
\label{almostmonogamy}
\end{lemma}
\begin{proof}[(outline)]
The lemma follows by using Eq. (\ref{DallequalsER}) and further developing the connection with hypothesis testing in the form of a new monogamy-like inequality for $E_R^{\infty}$:
\begin{eqnarray*} 
E_{R}^{\infty}(\rho_{A:BE}) - E_{R}^{\infty}(\rho_{A:E}) &=&  D_{\ALL}(\rho_{A:BE}) -  D_{\ALL}(\rho_{A:E})\\
 &\geq&  D_{\leftLOCC}(\rho_{A:B}).
\end{eqnarray*}
This inequality is proved by using measurements that achieve $D_{\ALL}(\rho_{A:E})$ and $D_{\leftLOCC}(\rho_{A:B})$ to construct a global measurement distinguishing $\rho_{ABE}$ from separable states $\CS_{A:BE}$ at a sufficiently good rate.
\end{proof}

We define in analogy to the LOCC norm, the one-way LOCC norm $\Vert * \Vert_{\leftLOCC}$, in which only measurements implementable by $\leftLOCC$ are allowed. Then next step is to convert the entropic bound on $I(A;B|E)$ obtained from Lemmas~\ref{nonlockability} and~\ref{almostmonogamy} into a lower bound in terms of the minimum $\leftLOCC$ distance to the set of separable states:
\begin{lemma}  For every $\rho_{AB}$, 
\[D_{\leftLOCC}(\rho_{A:B}) \geq \frac{1}{8 \ln 2}\Norm{\rho_{A:B} - \CS_{A:B}}_{\leftLOCC}^{2}.\]
\label{lowerboundnorm}
\end{lemma}
\begin{proof}[(outline)]
This follows from a combination of von Neumann's minimax theorem and Azuma's inequality, since separable states satisfy a martingale property when they are subject to local measurements. 
\end{proof}

So far, Lemmas 1,2 and 3 combine to give 
\begin{eqnarray*}
I(A;B|E) &{\geq}& E_R^\infty(\rho_{A:BE}) -  E_R^\infty(\rho_{A:E})\\
&{\geq}& D_{\leftLOCC}(\rho_{A:B}) \\
&{\geq}&  \frac{1}{8 \ln 2}\Norm{\rho_{A:B} - \CS_{A:B}}_{\leftLOCC}^{2}. 
\end{eqnarray*}
We now consider the family of norms $\norm{*}_{\LOCC(k)}$, which quantify distinguishability with respect to measurements that can be implemented by $k$ rounds of LOCC.  In particular, they satisfy $\leftLOCC = \LOCC(1)$ and $\LOCC = \cup_k \LOCC(k)$. 
Theorem~1 follows by recursive application of the following technical lemma, which is proved in \cite{BCY10}:
\begin{lemma}
Assume that 
\[I(A;B|E) \geq \frac{1}{8 \ln 2}\Norm{\rho_{A:B} - \CS_{A:B}}_{\LOCC(k-1)}^{2}.\]
Then 
\[I(A;B|E) \geq \frac{1}{8 \ln 2}\Norm{\rho_{A:B} - \CS_{A:B}}_{\LOCC(k)}^{2}.\]
\label{recursion}
\end{lemma}
\end{proof}

\section{Acknowledgments} We thank S. Aaronson, M. Berta, A. Harrow, L. Ioannou and A. Winter for helpful discussions.  FB and JY thank the Institute Mittag Leffler, where part of this work was done, for their hospitality.

\bibliographystyle{abbrv}

\end{document}